\documentstyle{article}
\begin{document}

\title{Cosmic Defects}

\author{Alexander Vilenkin\\
Institute of Cosmology, Physics Department,\\ 
Tufts University, Medford, MA 02155, USA}

\maketitle

\section{Defect zoo}

Many interesting developments in cosmology over the last 20 years have
come
from the realization, due to Kirzhnits \cite{Kirzhnits}, that (almost)
each
spontaneous symmetry breaking in particle physics corresponds to a
phase transition in the early universe.  For a typical grand-unified
theory, the early
moments after the big bang are characterized by the full grand unified
symmetry.  The exact number of phase transitions that break this
symmetry down to $SU(3)\times U(1)_{em}$ is model-dependent, but one
can expect at least two.  One at the energy scale of 
about $10^{16}$GeV where the strong
interaction became distinct from the electroweak interaction, and
another at 100 GeV where the electroweak symmetry was broken.  Of
course, there could be additional phase transitions at intermediate
scales.

Just like phase transitions in more familiar solids and liquids,
cosmological phase transitions can give rise to defects of various
kinds.  The defects are formed, roughly speaking, because the
directions of symmetry breaking are different in different regions of
space.  When these regions try to match at the boundaries, they
sometimes run into topological problems, and as a result we get
defects which trap the high-energy symmetric vacuum in their cores.  
The spatial variation in the directions of symmetry breaking
is inevitable due to causality.  At cosmic time $t$, correlations
cannot extend beyond the horizon distance, $d_H \sim ct$, since that
would require superluminal signal propagation.  The characteristic
correlation length, $\xi_c \leq ct$, which depends on the dynamics of
the phase transition, determines the typical distance
between defects.  This mechanism of
defect formation was first discussed by Kibble \cite{Kibble}.

Depending on the topology of the symmetry groups involved, the defects
can be in the form of surfaces, lines, or points.  They are called
domain walls, strings, and monopoles, respectively.  All three types
of defects are stable, in the sense that domain walls cannot develop
holes, strings cannot break, and monopoles cannot decay into other
particles.  This is guaranteed by topology and is independent of
the details of the models.

In addition to these elementary defects, the cosmic zoo includes
hybrid animals:  monopoles connected by strings, and domain walls
bounded by strings.  These can be formed in a sequence of phase
transitions, e.g., the first transition produces monopoles, which get
connected by strings at the second phase transition.  In some models
each monopole gets attached to $N$ strings.  For $N\geq 3$ this
results in the formation of a monopole-string network.

The defect zoo also includes textures, which do not have localized
cores and create a small region of high-energy vacuum only for a brief
instance of time, in the process of the unwinding of a topologically
non-trivial field configuration.

The physical properties of defects can be very different depending on
whether they are formed as a result of gauge or global symmetry
breaking.  Global defects have long-range Goldstone fields; the energy
density of these fields decreases rather
slowly with the distance, so that much of the defect energy is
distributed outside the core.  For gauge defects, the energy is very
well localized, and such defects can be well approximated as points,
lines, or surfaces of vanishing thickness.
  
A tremendous amount of research has been done on the formation,
evolution, and cosmological consequences of various defects.  
One finds that domain walls and monopoles are disastrous
for cosmological models and should be avoided \cite{footnote1}.  The
simplest hybrid defects, monopoles connected by strings and domain
walls bounded by strings, are not dangerous, but they decay soon after
they are formed.  The remaining defects may exist in the present
universe and can produce potentially detectable observational effects.
All types of defects, including the `dangerous' and transient ones,
have been proposed for one cosmological role or another.

For a review of topological defects and their cosmological
implications, the reader is referred to \cite{Book}.  The literature
on this subject is rapidly expanding, and below I give only scattered
references to some work that appeared after the publication of
Ref. \cite{Book}.

\section{Cosmic roles for defects}

Much of the research on topological defects was motivated by defect
models of structure formation.  Gauge strings and global strings,
monopoles and textures have all been suggested as possible seeds of
galaxies and large-scale structure.  Only superheavy  defects, with a
grand-unification scale of symmetry breaking, are suitable for this
role.  An attractive feature of this class of structure formation
scenarios is that they are directly verifiable: peculiar gravitational
interactions of defects should allow one to detect their presence in
the universe today.  The most promising observational probe appears to
be the pattern of temperature distribution of the microwave background
on small angular scales.  All defects introduce non-Gaussian features
in this pattern, distinguishing them from one another and from
the competing models based on inflationary scenarios \cite{MB}.

Superheavy cosmic strings can also produce multiple images of distant
galaxies and clusters and can generate an observable gravitational-wave
background ranging over many decades in frequency.  Hybrid defects,
which will have decayed long before the present time, 
may still leave a characteristic
signature in this background \cite{MV}.  
Global defects create a background of
massless Goldstone bosons.  In axion models, where strings are
produced due to an approximate global symmetry breaking, the resulting
pseudo-Goldstone bosons (axions) have a small mass and are prime
candidates for cold dark matter.

If the constituent fields of the defects have baryon-number-violating
interactions, their decay could result in the generation of the
observed baryon asymmetry of the universe.  `Regular' cosmic strings, 
monopoles connected by strings, and monopole-string networks have been
suggested for this role.

Topological defects can produce high-energy particles by a variety of
mechanisms and can contribute to the observed spectrum of cosmic rays.
A particularly intriguing possibility is that ultra-high energy 
cosmic rays with $E\geq 10^{11}$ GeV, which are hard to explain by the
standard Fermi acceleration mechanism, may be due to vacuum defects
\cite{HS}.  The prime suspects here are superconducting cosmic strings
and monopole-string networks.  I should also mention the idea that
cosmic rays are produced as a result of monopole-antimonopole
annihilation \cite{BS} 
and the astonishing proposal that the ultrahigh-energy
particles {\it are} magnetic monopoles \cite{WK}.

In this list of the proposed roles for the defects, I made no attempt
at completeness.  Different roles require different types of defects
with different energy scales and couplings to ordinary matter.  Which,
if any, of these proposals are true depends on the particle physics at
very high energies (of which we know very little).  Even if defects
played no prominent cosmological roles, the search for their
observational signatures is still very much worth pursuing.  Needless
to say, if topological defects are discovered, we are likely to learn
a great deal both about particle physics and early universe cosmology.

\section{Defects and inflation}

Inflationary cosmological models explain the homogeneity, isotropy and
flatness of the universe.  Since no competing theories that
can explain these facts have surfaced in the 15 years since inflation
was first proposed, we seem to have little choice but to assume that
the early universe did go through a period of inflation.  Quantum
fluctuations during this period {\it could} create cosmologically
significant density fluctuations.  But it is quite possible that they
did not, in which case the observed structures could be seeded by
topological defects.

Topological defects which formed before inflation would have been
drastically diluted by the expansion and never seen again.  On the
other hand, the thermalization temperature of the universe after
inflation is unlikely to exceed $10^{16}~GeV$, and it is often said
that 
superheavy defects needed for structure formation are incompatible
with inflation.  This, however, is far from being true: in a wide
class of particle physics models, defects and inflation can peacefully
coexist with one another.

Here are some of the possibilities.  (i) The symmetry-breaking phase 
transition could occur during inflation, but sufficiently close to its
end, so that the defects are not completely inflated away.  
Such phase transitions are
driven not by the temperature, but by the curvature or by the evolving
inflaton field.  If defects are formed within 30 e-foldings of the end
of inflation, they can generate density fluctuations on galactic and
larger scales.  (ii) In models of `extended' inflation, superheavy 
defects can be produced in bubble collisions at the end of inflation.
(iii) Defects can be produced in a `pre-heating' transition after
inflation.  The amplitude of scalar-field fluctuations at pre-heating
can reach Planckian values, and superheavy defects can be formed even
if the eventual thermalization temperature is very low \cite{LKS}.  
(iv) In some
supersymmetric models, certain couplings can be naturally small, and
superheavy strings can be formed as late as the electroweak phase
transition.  None of these options requires any drastic fine-tuning
of the parameters, and some of them require none.

\section{Concluding remarks}

In conclusion, I would like to emphasize that formation of topological
defects at a phase transition is a very generic phenomenon.  Since the
early universe probably went through several phase transitions, it is
rather unlikely that no defects at all were formed.  A discovery of
topological defects would open a direct window into the physics of
very high energies.  And I think there is a reasonable chance that
this will actually happen in not so distant future.


\begin{thebibliography}{}
\bibitem{footnote1}
Exceptions to
this rule are models with very light domain walls of symmetry breaking
scale below 1 MeV and inflationary models where a small density of
monopoles is produced near the end of inflation.
\end{thebibliography}
\end{document}